\documentclass[twocolumn,showpacs,10pt,prb,floatfix,amssymb]{revtex4-1}

\usepackage{graphicx}

\begin{document}

\newcommand{\nx}{\textrm}

\title{Transport properties of monolayer MoS$_2$ grown by chemical vapour deposition}
\author{H. Schmidt$^{+~1,2}$}
\author{S. Wang$^{1,2}$}
\author{L. Chu$^{1,2}$}
\author{M. Toh$^{1,2}$}
\author{R. Kumar$^{1,2}$}
\author{W. Zhao$^{1,2}$}
\author{A.~H.~Castro Neto$^{1,2}$}
\author{J. Martin$^{1,2}$}
\author{S. Adam$^{1,2,4}$}
\author{B. Oezyilmaz$^{1,2}$}
\author{G. Eda$^{*~1,2,3}$}

\affiliation{$^1$Graphene Research Centre, National University of Singapore, 6 Science Drive 2, Singapore 117546\\
$^2$Department of Physics, National University of Singapore, 2 Science Drive 3, Singapore 117542\\
$^3$Department of Chemistry, National University of Singapore, 3 Science Drive 3, Singapore 117543\\
$^4$Yale-NUS College, 6 College Avenue East, Singapore, 138614}

\date{\today}
\begin{abstract}
Recent success in the growth of monolayer MoS$_2$ via chemical vapor deposition (CVD) has opened up prospects for the implementation of these materials into thin film electronic and optoelectronic devices. Here, we investigate the electronic transport properties of individual crystallites of high quality CVD-grown monolayer MoS$_2$. The devices show low temperature mobilities up to 500~cm$^2$V$^{-1}$s$^{-1}$ and a clear signature of metallic conduction at high doping densities. These characteristics are comparable to the electronic properties of the best mechanically exfoliated monolayers in literature, verifying the high electronic quality of the CVD-grown materials. We analyze the different scattering mechanisms and show, that the short-range scattering plays a dominant role in the highly conducting regime at low temperatures. Additionally, the influence of phonons as a limiting factor of these devices is discussed.
\end{abstract}
\maketitle
Two-dimensional (2D) crystals of transition metal dichalcogenides (TMD) have received significant interest due to their potential in a wide range of novel applications as well as in basic research \cite{r1,r2,r3}. Especially monolayers of semiconducting TMDs such as MoS$_2$ hold significant promise in electronics and optoelectronics due to their unusual electrostatic coupling \cite{r4}, large carrier mobility \cite{r5}, high current carrying capacity \cite{r6}, and strong absorption in the visible frequencies \cite{r7}, on top of their chemical and mechanical robustness. Strong spin-orbit coupling and the unique crystal symmetry of these materials lead to the coupling of spin and valley degrees of freedom, which can be exploited for the development of novel valleytronic devices \cite{r8}.

\begin{figure}[ht!]
\includegraphics[width=0.84\columnwidth]{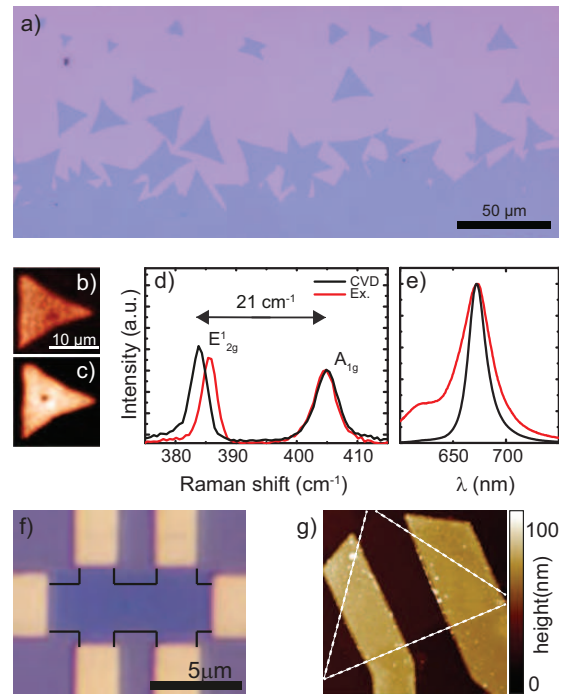}
\caption{\label{fig:bild1} a) Optical image of CVD grown monolayer MoS$_2$ on silicon substrate with a continuous film in the bottom and single triangles above. b) Intensity mapping of the Raman E$_1$ peak of one crystal. The spectrum at one point is shown in (d), with a distance of 21 cm$^{-1}$ between two vibrating modes (E$^{1}_{2g}$ and A$_{1g}$), characteristic for CVD monolayer MoS$_2$. The spectrum for an exfoliated monolayer exited with the same parameters is shown in red. c) Photo luminescence intensity map of the flake and e) corresponding normalized spectrum, again compared to an exfoliated sample (red). f) Optical picture of a structured and contacted Hall-bar used for multiterminal measurements. g) Atomic force microscopy image (10~$\mu m$) of a two terminal device. The shape of the triangular crystal is highlighted.}
\end{figure}

While monolayers of MoS$_2$ can be readily obtained by micromechanical cleavage of synthetic or natural bulk crystals \cite{r1}, large area, high quality, and continuous thin films are needed for practical devices. To this end, several groups have recently succeeded in the growth of monolayer thin films TMDs via chemical vapor deposition (CVD) \cite{r9,r10,r11,r12,r13}. The formation of MoS$_2$ monolayers during the CVD process occurs through nucleation and growth, resulting in a film which consists of misoriented grains stitched together by lines of 8- and 4-membered rings \cite{r10,r11}. Initial studies have suggested that the grain boundaries have minor effects on the charge transport properties \cite{r10}. The reported room temperature carrier mobility of these thin films, which is often used as a measure of the electronic quality, is found to be between 0.1 and 10~cm$^2$V$^{-1}$s$^{-1}$ for unencapsulated devices. These values are almost in the same order of magnitude compared to those measured from mechanically exfoliated counterparts but distinctly lower than the theoretically predicted values \cite{r14}. While recent transport studies suggest that charged impurities \cite{r15,r16,r17} and localized states \cite{r16,r18,r19} play a crucial role for mechanically exfoliated samples, the dominant scattering processes that limit the carrier mobility remains elusive.\\

In this article we report on the electronic transport properties of CVD-grown crystallites of monolayer MoS$_2$ and demonstrate that their electronic quality is comparable to that of mechanically exfoliated materials. In back-gated device geometry without encapsulation, our CVD MoS$_2$ monolayers exhibit room and low temperature field effect mobilities of up to 45 and 500~cm$^2$V$^{-1}$s$^{-1}$, respectively. We also report the observation of crossover from insulating to metallic conduction as a function of carrier density and temperature. This phenomenon, previously referred to as metal-insulator-transition (MIT) \cite{r15}, allows access to the transport regime where the effect of band edge disorder can be neglected. Our analysis shows that carrier mobility in the high charge carrier density regime is largely limited by structural defects at low temperatures.\\

Our CVD MoS$_2$ thin films were grown on silicon substrates covered with 300 nm of silicon dioxide using a method reported by other groups \cite{r9, r10}. Near the edge of the continuous MoS$_2$ film, numerous large ($>$ 10~$\mu m$) crystallites of monolayer MoS$_2$ are found (Figure 1a). The triangular shape of most crystallites reflects the three-fold symmetry of MoS$_2$ suggesting they are single-crystalline. Uniform Raman and photoluminescence signals (Figure 1b and c) from most individual crystallites further verify that they consist of a single crystal domain with no internal grain boundaries \cite{r10}. Note that some triangular crystallites consisting of multiple domains were occasionally found in different batches (See Supplementary Information for details). The sharp spectral features indicate evidence of no substantial disorder in the sample (Figure 1d and e). In fact, the band gap photoluminescence from the CVD samples shows a distinctly sharper peak compared to that from mechanically exfoliated counterparts, reflecting their high electronic quality. In the following, we discuss the transport properties of CVD-grown MoS$_2$ based on devices fabricated from individual crystallites (See Methods for details of the device fabrication). Several devices in both two terminal (Fig 1g) and multi-terminal geometry (Fig 1f) were studied.\\

\begin{figure}[ht!]
\includegraphics[width=0.95\columnwidth]{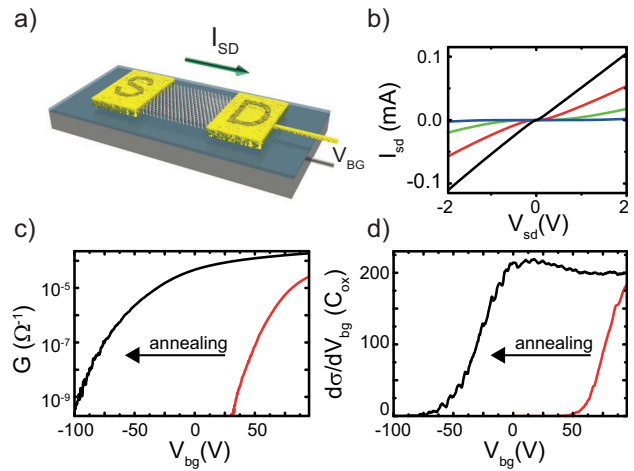}
\caption{\label{fig:bild2} a) Sketch of the field effect device. A monolayer of MoS$_2$ is contacted with gold electrodes on a Si/SiO$_2$ substrate, on which a backgate voltage $V_{bg}$ is applied to tune the Fermi energy of the sample. b) I-V curves of two terminal measurements before in situ annealing at different gate voltages of 25, 45, 65, and 85 V (black) and a temperature of 5 K. c) Gate dependence of the conductance before (red) and after (black) in-situ annealing, measured at 10 K. d) Differential conductivity in units of the gate capacitance, equivalent to the field effect mobility before and after annealing.}
\end{figure}

After growth, the CVD MoS$_2$ film and triangles are transferred to a fresh silicon substrate with thermally grown oxide which is used as the backgate dielectric. Selected devices are then contacted using standard electron beam lithography (Figure 2a). The output characteristic of a typical two-terminal device (Figure 2b) indicates good ohmic contact with the gold electrodes at large positive gate biases even at low temperatures \cite{r4,r17}.  In the highly conducting regime, the contact resistance plays a minor role and the 2-probe and 4-probe measurements yield similar results \cite{r5}. The activating behavior observed at lower gate biases indicates the insulating regime of the MoS$_2$ and the non-negligible effects of contact resistance due to Schottky barriers. In order to achieve optimal device performance, we employed a two-step annealing process \cite{r5}: first at 200~$^\circ$C for 2 hours in N$_2$ and subsequent annealing in vacuum at 120~$^\circ$C for 4-10 hours. Transport measurements were performed immediately after the second annealing step without exposing the device to ambient. This second annealing had a significant effect on the doping level as evidenced by the significant shift in the transfer curve towards the negative gate bias, making the device more strongly n-type (Figure 2c). This is likely due to the removal of adsorbents such as O$_2$ or H$_2$O, which are known to deplete negative charge carriers \cite{r20} and the resulting shift of the Fermi level towards the conduction band. It is worth noting that the threshold shift due to vacuum annealing can be as large as 100~V, which corresponds to increase in the carrier concentration by 7$\times$10$^{12}$ cm$^{-2}$. As discussed below, this allows us to readily access the metallic conduction regime with relatively small backgate voltages without the use of ionic \cite{r21} or high-κ dielectric topgate \cite{r15}. The channel resistivity was found to be below 10 k$\Omega$, which is among the lowest values reported to date for monolayer MoS$_2$. At large negative gate biases, the insulating regime is achieved yielding on/off ratios of $>$10$^6$ at low temperatures.\\

In backgated devices with no dielectric encapsulation, field-effect mobility $\mu_{fe}$ can be obtained by $\mu_{fe} = d\sigma$/dV$_{bg} * $C$_{ox}^{-1}$ where C$_{ox}$=11.5 nF is the gate oxide capacitance, $\sigma$ is the channel conductivity, and V$_{bg}$ is the backgate bias. Figure 2d shows the gate bias dependence of the field-effect mobility of a two-probe device at 10 K before and after the vacuum annealing step. The maximum $\mu_{fe}$ of 200~cm$^2$V$^{-1}$s$^{-1}$ remains almost unchanged but the saturation of mobility, which corresponds to the linear regime of the conductivity, is observed only after the second annealing step. It may be noted that both 2- and 4-probe field effect mobilities were found to be in a range between 100 to 500~cm$^2$V$^{-1}$s$^{-1}$  and the discrepancies between the two were only evident near the insulating regime (where R $>$ 10 M$\Omega$) where the contact resistance becomes significant \cite{r5}. Due to large uncertainties in the Hall mobilities and the large range in resistances studied, we focus the following discussions on field effect mobilities, which are obtained from DC two-terminal measurements.\\

\begin{figure}[t]
\includegraphics[width=0.95\columnwidth]{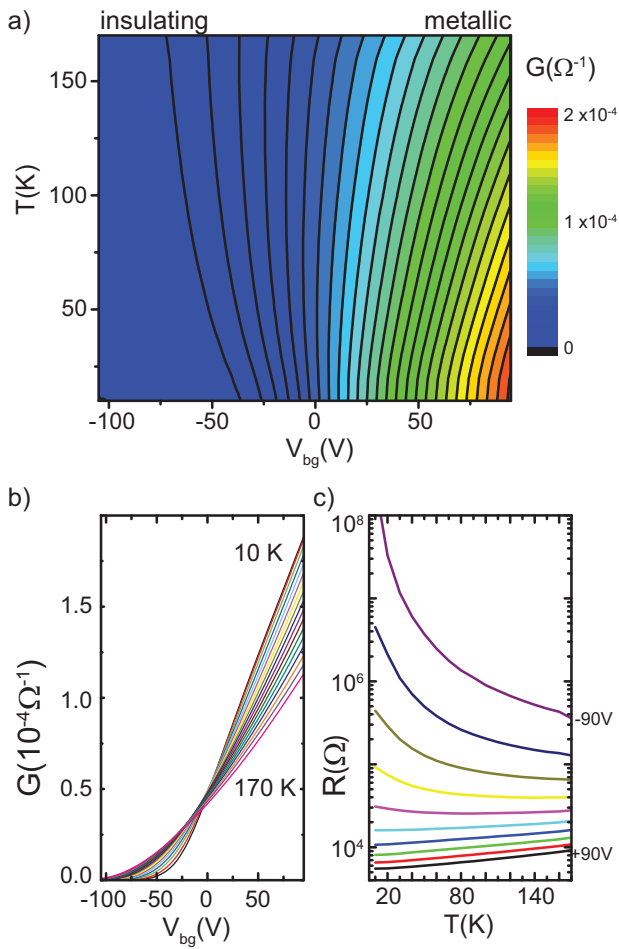}
\caption{\label{fig:bild3} a) Color plot of the conductance as a function of temperature and gate voltage. b) Conductance as a function of gate voltage for different temperatures. The crossing around $V_{bg}$=0~V indicates the change in temperature dependence. c) Temperature dependence of the 2-terminal resistance at different gate voltages from 90 to -90~V.}
\end{figure}

Figure 3a shows the conductance of a device as a function of backgate voltage and temperature. For positive gate voltages, the conductance decreases with increasing temperature, indicating metallic behavior, whereas for negative biases the temperature dependence is reversed, showing the characteristic of an insulator. This crossover from insulating to metallic conduction is shown in more detail in Figures 3b and c. The transfer curves show a gradually shifting crossover point around V$_{bg}$ = 0~V above and below which the temperature dependence is opposite. The crossover point occurs at the channel conductivity in the order of e$^2$/h similar to the previous reports by Radisavljevic and Kis \cite{r15} and Baugher et al. \cite{r5}, suggesting that they share the same origin. In most of our devices, the crossover point occurs at low gate voltages, allowing the analysis of carrier conduction in the fully metallic regime where band edge disorder plays a minor role. The resistance shows an increase with temperature in the metallic regime while it decreases in the insulating regime, suggesting phonon-assisted hopping conduction \cite{r16,r18} and thermal activation of carriers.\\

The low temperature field effect mobility of our devices is among the highest reported to date, however, it falls significantly below the acoustic-phonon-limited mobility of $10^5$~cm$^2$V$^{-1}$s$^{-1}$ predicted by theory \cite{r14}. This strongly suggests that the mobility is limited by extrinsic factors at low temperatures. The gate-dependent field effect mobilities shown for different devices in Figures 4a and b can be used to distinguish three different transport regimes. For large negative voltages, transport is dominated by variable range hopping (See supplementary information for details), as reported previously \cite{r16, r19}. On the other hand, at lower gate biases around zero, the field effect mobility increases with gate voltage. This behavior corresponds to $\sigma\propto n^\alpha$ with $1<\alpha<2$, which is indicative of charged impurity scattering in a two dimensional system with parabolic dispersion \cite{r22} and T $>$ T$_{Fermi}$. Since the material is strongly doped by electron donors, the presence of ionized atoms or Coulomb impurities is expected. For high positive voltages the system is in the metallic regime and reveals an upper limit of mobility, which is density-independent at low temperatures ($<$ 40 K) as also shown in Fig.~4b. This saturation of mobility with temperature as well as with charge carrier density is a signature of short-range scattering limiting the device performance in the highly conductive state. It is worth noting that a similar behavior was observed only for topgated devices with high-$\kappa$ dielectric \cite{r15} also reaching high charge carrier densities.\\

As the temperature increases, phonon scattering due to acoustic and polar optical phonons are expected to play a dominant role in the metallic regime and decrease mobility.  In the presence of more than one scattering mechanism, Matthiessen’s rule can be used to describe the contributions from the various scatterers:
\begin{equation}
\frac{1}{\mu_{total}}=\frac{1}{\mu_{ph}}+\frac{1}{\mu_{sr}}+\frac{1}{\mu_{lr}}
\end{equation}
where $\mu_{ph}$, $\mu_{sr}$, and $\mu_{lr}$ represent the mobilities limited by phonons, short-range, and long-range scatterers. Here we assume other contributions such as electron-electron scattering to be minor. At low temperatures, $\mu_{ph}$ can be neglected in our system because the other scattering mechanisms dominate. We also can neglect $\mu_{lr}$ in the high voltage regime since charged impurities give a super-linear conductivity and we have $\mu_{sr}$ $<<$ $\mu_{lr}$.  As shown in Fig 4a, b, we extract $\mu_{sr}$ to be of the order 200~cm$^2$V$^{-1}$s$^{-1}$ for this sample.\\

In the metallic conduction regime, the phonon contribution leads to mobility damping with a power law dependence on temperature $\mu_{ph}\propto $T$^{-\gamma}$. The theoretical analysis by Kaasbjerg et al. \cite{r14} predicts $\gamma$ to vary from 1 (at 100~K) to 1.7 (at room temperature). At very low temperatures when only acoustic phonons are the limiting factor, $\gamma$ approaches 1, but above ~100 K optical phonons play a dominant role and increasing $\gamma$ is expected. The reported damping factor $\gamma$ obtained from transport experiments on monolayer MoS$_2$ varies between 0.62-1.7  for unencapsulated devices\cite{r5, r15, r16} and shows distinctly lower values of 0.3 - 0.78 for devices with high-κ top-gate dielectric \cite{r5}. While some variations in the apparent phonon damping factor may be explained by charged impurity scattering \cite{r3} and homopolar phonon quenching \cite{r14}, the origin of the observed variations is unclear. We show below that multiple factors affect the apparent phonon damping factor and extraction of the true phonon contribution requires careful analysis.\\

\begin{figure}[ht!]
\includegraphics[width=0.95\columnwidth]{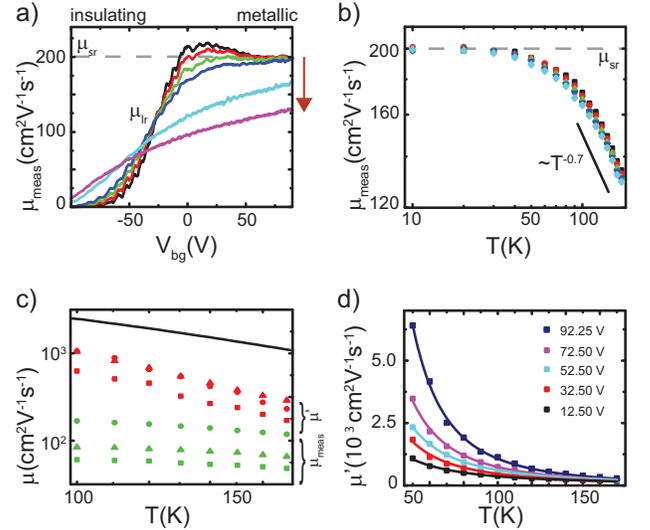}
\caption{\label{fig:bild4} a) Mobility as a function of gate voltage. Temperatures are 10, 20, 30, 40, 100, and 150 K. The regions limited by short and long range scatterers as well as the phonon influence (red arrow points towards higher temperatures) can be clearly distinguished. b) Temperature dependence of the measured mobility at different gate voltages from 67.5 to 87.5~V in double logarithmic scaling. The black line follows $T^{-0.7}$ as guide to the eyes. c) Mobilities for the phonon dominated part at temperatures above 100 K. The as-measured mobilities $\mu_{meas}$ for three devices are shown as well as the values of $\mu`$. The black line indicates the theoretical prediction for the mobility limited by phonon scattering obtained from literature \cite{r14}. d) The mobility $\mu`$ as a function of temperature for several gate voltages with the according power-law fits. Each curve has been averaged over multiple gate voltages (+- 1 V) before fitting. The plot shows that the $\gamma^’$ values strongly depend on the gate bias (See Supporting Information for details).}
\end{figure}

The common feature of the previously reported results is that the low temperature mobility is saturated in temperature around 100-200~cm$^2$V$^{-1}$s$^{-1}$ independent of the presence of high-$\kappa$ top-gate dielectric. Also most of these measurements do not reach the linear conductivity regime at high charge carrier densities, the saturation in temperature suggest that the mobility is predominantly limited by short-range scatterers as we observe in our samples. Once multiple scattering mechanisms are present, the damping factor can be appearing to be strongly suppressed from its intrinsic value. We found that the power-law fit of the as-measured mobilities below 170~K yields apparent low $\gamma$ values of 0.7 (Fig 4b). But correct analysis of the intrinsic damping factor requires that contributions to scattering due to phonons be separated from those due to other scattering mechanisms. Since $\mu_{sr}$ is known and temperature independent, we can subtract this contribution from the measured mobility to obtain $\mu`$ = (1/$\mu_{meas}$-1/$\mu_{sr}$)$^{-1}$. Figure 4c shows the as-measured $\mu_{meas}$ and calculated $\mu`$ at high gate voltage for three different devices on a double logarithmic plot. While both $\mu_{meas}$ and $\mu`$ can be fitted reasonably well with power law dependence, the damping factor is significantly higher for $\mu`$. We can apply the same analysis to the previously reported results \cite{r15} in the highly conductive regime and achieve similar increase in the $\gamma$ value (see Supplementary Information for details).\\

Figure 4d shows the temperature damping of $\mu`$ for several gate voltages with the according fits. The curves follow the expected $\propto T^{-\gamma `}$ very well in this temperature regime and the extracted values for the damping factor ’ vary with backgate voltage between 1 and 2.6. This points towards the collective role of multiple scattering mechanisms besides phonons in the metallic regime influencing the transport \cite{r23}. Large damping factors approaching that of the bulk ($\gamma_{bulk} =2.6)$\cite{r24} observed in our analysis suggest that the effects of optical phonons may be dominant even at low temperatures and could indicate that the deformation potentials for the acoustic and optical phonons in monolayer MoS$_2$ are considerably different from the values expected theoretically \cite{r14}.\\

In summary, we demonstrate that the electronic transport properties of CVD-grown monolayer MoS$_2$ samples are comparable to those of mechanically exfoliated samples. Despite the common perception that CVD-grown thin films are structurally defective due to thermal stresses caused during the growth process and morphological features introduced during the transfer process, large field effect mobility was achieved in non-optimized backgated geometry. Our observation of the crossover from the insulating to metallic conduction verifies the high electronic quality and inherent n-type doping of the material. We further show that short-range scatterers limit the mobility at high carrier densities at low temperatures, and discuss the influence of phonons at elevated temperatures. Our results provide positive prospects for further improvements of the device performance and device implementation of CVD MoS$_2$ thin films.\\
G.E and S.A. acknowledge Singapore National Research Foundation for funding the research under NRF Research Fellowship (NRF-NRFF2011-02, NRF-NRFF2012-01)
B.O. acknowledges the NRF-CRP award “Toward Commercialization of Graphene Technologies” (R-144-000-315-281).
A.H.C.N. acknowledges the NRF-CRP award 'Novel 2D materials with tailored properties: beyond graphene' (R-144-000-295-281).
\\\\
Corresponding Authors:\\
$^{+}$Hennrik Schmidt, Email: physche@nus.edu.sg\\
$^{*}$Goki Eda, Email:  g.eda@nus.edu.sg\\

\newpage

\begin{thebibliography}{10}
\bibitem{r1} K. S. Novoselov, D. Jiang, F. Schedin, T. J. Booth, V. V. Khotkevich, S. V. Morozov, and  A. K. Geim, {\em PNAS} {\bf 102}, 10451 (2005).

\bibitem{r2} Q. H. Wang, K. Kalantar-Zadeh, A. Kis, J. N. Coleman, and M. S. Strano, {\em Nature Nanotech.} {\bf 7}, 699 (2012).

\bibitem{r3} M. Chhowalla, H. S. Shin, G. Eda, L.-J. Li, K. P. Loh, and Hua Zhang, {\em Nat. Chem.} {\bf 5}, 263 (2013).

\bibitem{r4} B. Radisavljevic, A. Radenovic, J. Brivio, V. Giacometti, and A. Kis, {\em Nature Nanotech.} {\bf 6}, 147, (2011).

\bibitem{r5} B. W. H. Baugher, H. O. H. Churchill, Y. Yang, and P. Jarillo-Herrero, {\em Nano Lett.} {\bf 13}, 4212, (2013).

\bibitem{r6} D. Lembke and A. Kis, {\em ACS Nano} {\bf 6}, 10070 (2012).

\bibitem{r7} G. Eda, and S. A. Maier, {\em ACS Nano} {\bf 7}, 5660 (2013). 

\bibitem{r8} D. Xiao, G.-B. Liu, W. Feng, X. Xu, and W. Yao, {\em Phys. Rev. Lett.} {\bf 108}, 196802 (2012).

\bibitem{r9} Y.-H. Lee, X.-Q. Zhang, W. Zhang, M.-T. Chang, C.-T. Lin, K.-D. Chang, Y.-C. Yu, J. T.-W. Wang, C.-S. Chang, L.-J. Li, and T.-W. Lin, {\em Advanced Materials} {\bf 24}, 2320 (2012).

\bibitem{r10} A. M. van der Zande, P. Y. Huang, D. A. Chenet, T. C. Berkelbach, Y. You, G.-H. Lee, T. F. Heinz, D. R. Reichman, D. A. Muller, and J. C. Hone, {\em Nature Materials} {\bf 12}, 554 (2013).

\bibitem{r11} S. Najmaei, Z. Liu, W. Zhou, X. Zou, G. Shi, S. Lei, B. I. Yakobson, J.-C. Idrobo, P. M. Ajayan, and J. Lou, {\em Nature Materials} {\bf 12}, 754 (2013). 

\bibitem{r12} H. Wang, L. Yu Y.-H. Lee, W. Fang, A. Hsu, P. Herring, M. Chin, M. Dubey, L.-J. Li, J. Kong, and T. Palacios, {\em Electron Devices Meeting, 2012 IEEE International} 4.6.1 (2012).

\bibitem{r13} W. Wu, D. De, S.-C. Chang, Y. Wang, H. Peng, J. Bao, and S.-S. Pei, {\em Appl. Phys. Lett.} {\bf 102}, 142106 (2013).

\bibitem{r14} K. Kaasbjerg, K. S. Thygesen, and K.W. Jacobsen, {\em Phys. Rev. B} {\bf 85}, 115317 (2012).

\bibitem{r15} B. Radisavljevic, and A. Kis, {\em Nature Materials} {\bf 12}, 815 (2013). 

\bibitem{r16} D. Jariwala, V. K. Sangwan, D. J. Late, J. E. Johns, V. P. Dravid, T. J. Marks, L. J. Lauhon, and M. C. Hersam, {\em Appl. Phys. Lett.} {\bf 102}, 173107 (2013).

\bibitem{r17} S.-L. Li, K. Wakabayashi, Y. Xu, S. Nakaharai, K. Komatsu, W.-W. Li, Y.-F. Lin, A. Aparecido-Ferreira, and K.Tsukagoshi, {\em Nano Lett.} {\bf 13}, 3552 (2013).

\bibitem{r18} H. Qiu, T. Xu, Z. Wang, W. Ren, H. Nan, Z. Ni, Q. Chen, S. Yuan, F. Miao, F. Song, G. Long, Y. Shi, L. Sun, J. Wang, and  X. Wang, {\em Nature Comm.} {\bf 4}, 2642 (2013).

\bibitem{r19} S. Ghatak, A. N. Pal, and A. Gosh, {\em ACSNano} {\bf 5}, 7707 (2011).

\bibitem{r20} Sefaattin Tongay, J. Zhou, C. Ataca, J. Liu, J. S. Kang, T. S. Matthews, L. You, J. Li, J. C. Grossman, and J. Wu, {\em Nano Lett.} {\bf 13}, 2831 (2013).

\bibitem{r21} M. M. Perera, M.-W. Lin, H.-J. Chuang, B. P. Chamlagain, C. Wang, X. Tan, M. M.-C. Cheng, D. Tomnek, and Z. Zhou, {\em ACS Nano} {\bf 7}, 4449 (2013).

\bibitem{r22} S. Das Sarma, S. Adam , E. H. Hwang , and E. Rossi, {\em Rev. Mod. Phys.} {\bf 83}, 407 (2011).  

\bibitem{r23} Z.-Y. Ong and M. V. Fischetti, {\em Phys. Rev. B} {\bf 88}, 165316 (2013).

\bibitem{r24} R. Fivaz and E. Mooser, {\em Phys. Rev.} {\bf 163}, 743 (1967).








\end{thebibliography}
\end{document}


\newcommand{\nx}{\textrm}

\title{Supplementary Information for:\\Transport properties of monolayer MoS$_2$ grown by chemical vapour deposition}
\author{H. Schmidt$^{+~1,2}$}
\author{S. Wang$^{1,2}$}
\author{L. Chu$^{1,2}$}
\author{M. Toh$^{1,2}$}
\author{R. Kumar$^{1,2}$}
\author{W. Zhao$^{1,2}$}
\author{A.~H.~Castro Neto$^{1,2}$}
\author{J. Martin$^{1,2}$}
\author{S. Adam$^{1,2,4}$}
\author{B. Oezyilmaz$^{1,2}$}
\author{G. Eda$^{*~1,2,3}$}

\affiliation{$^1$Graphene Research Centre, National University of Singapore, 6 Science Drive 2, Singapore 117546\\
$^2$Department of Physics, National University of Singapore, 2 Science Drive 3, Singapore 117542\\
$^3$Department of Chemistry, National University of Singapore, 3 Science Drive 3, Singapore 117543\\
$^4$Yale-NUS College, 6 College Avenue East, Singapore, 138614}

\date{\today}
\maketitle
\section{Sample Preparation}
Due to the high temperatures during growth, the underlying silicon dioxide becomes defective and therefore loses its function as an insulating layer for backgating. To solve this issue, the samples are transferred to another Si/SiO$_2$ substrate using PMMA as transfer film and a KOH etch to remove the silicon dioxide. After the samples are transferred to a new wafer, they are cleaned carefully. Some devices have been structured by oxygen plasma or XeF$_2$ etching. Contacts are evaporated in a two-step process. While almost all parts consist of Cr/Au, the last small tips actually contacting the MoS$_2$ are made with pure gold (50 nm), which has proven to yield better contacts.
\section{Optical analysis}
Raman and photoluminescence (PL) spectra were obtained by a confocal Raman spectrometer in the back scattering geometry with a 532~nm excitation laser. The intensity is kept at 30~$\mu$W for all measurement in order to avoid any sample damage or local heating. The reference sample was mechanically exfoliated from natural MoS$_2$ crystals (SPI supplies). For the Raman measurement, a 2400 grooves per mm grating was used to obtain high spectral resolution of about 0.5~cm$^{-1}$. The Raman spectra are normalized by the intensity of the Raman mode from silicon. The PL spectra for CVD grown and exfoliated MoS$_2$ are obtained under the same measurement condition in order to directly compare their intensity.
\begin{figure}[t]
\includegraphics[width=0.68\columnwidth]{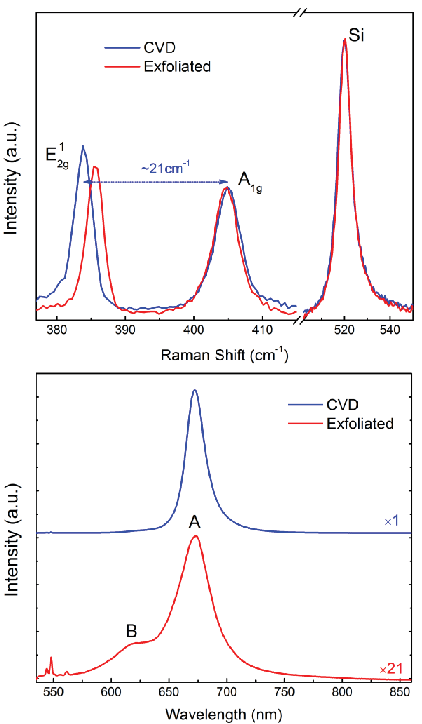}
\caption{\label{fig:bild1} Raman and photoluminescence spectra of CVD (blue) and exfoliated (red) monolayer MoS$_2$ }
\end{figure}\\
The frequency difference between E$_{2g}^1$  and A$_{1g}$  modes for exfoliated monolayer MoS$_2$ is around 19.5 cm$^{-1}$, agreeing well with other reports \cite{s1}. In the case of the grown MoS$_2$ it is about 21~cm$^{-1}$, which is a typical value for CVD grown MoS$_2$ \cite{s1,s2}. Two prominent PL peaks, marked with A and B, from exciton A and B at K point of the Brillouin zone are observed for exfoliated monolayer MoS$_2$. In contrast to this, the A peak with much higher intensity and narrower width is dominant in the PL spectrum of the CVD grown monolayer MoS$_2$. It is well established that the intrinsic doping and defect in exfoliated MoS$_2$ greatly suppress its PL quantum yield \cite{s1,s3,s4}. This strong difference could therefore indicate less intrinsic doping and defect, i.e. higher crystal quality, for the CVD grown MoS$_2$. In addition, the uniform PL and Raman mapping also indicates the high crystal quality of the CVD grown MoS$_2$. Nevertheless, some of the triangular crystals exhibit internal grain boundaries in the PL as shown in Figure 2.
\begin{figure}[ht]
\includegraphics[width=0.27\columnwidth]{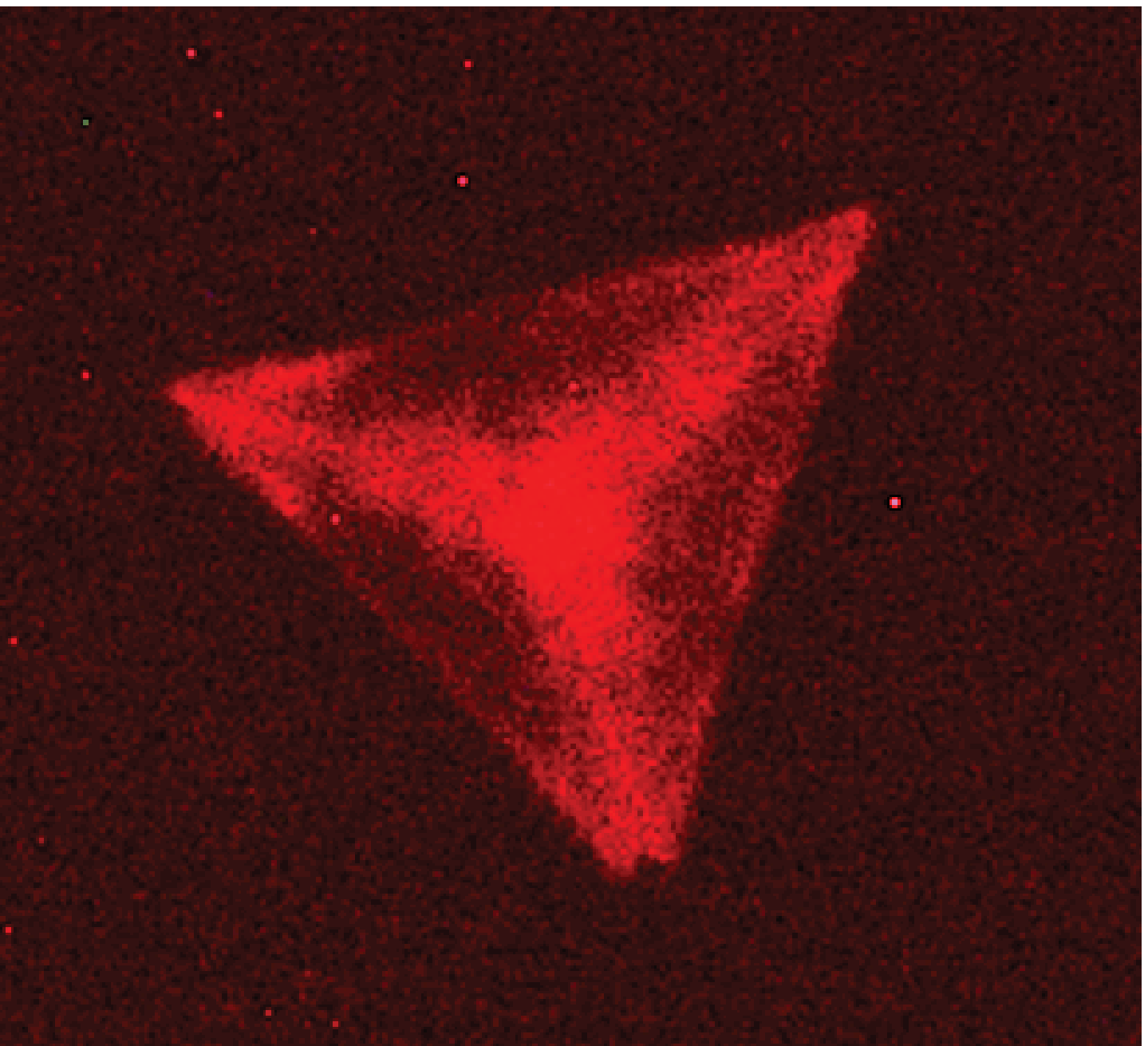}
\caption{\label{fig:bild2} Photoluminescence picture of one triangular sample, showing the internal structure. }
\end{figure}
\newpage
\section{Annealing}
\begin{figure}[th]
\includegraphics[width=0.95\columnwidth]{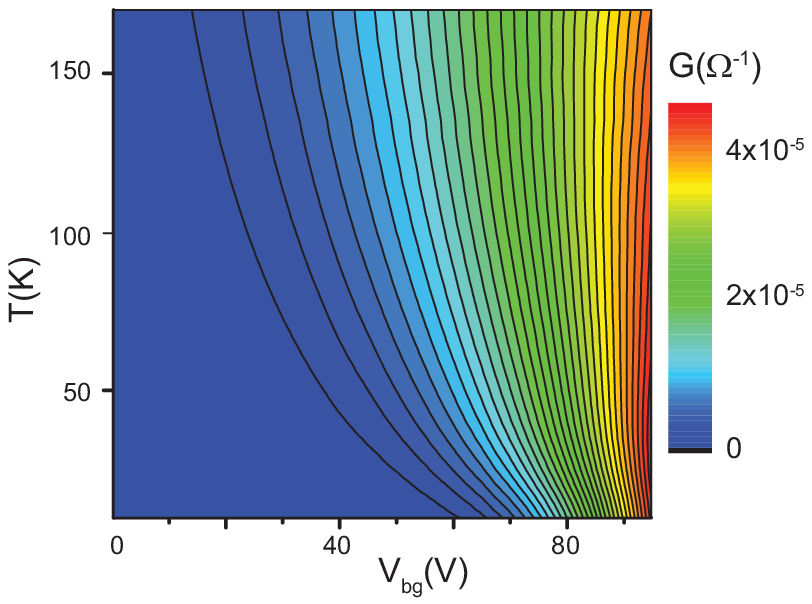}
\caption{\label{fig:bild3} Conductance of a device before in-situ annealing.}
\end{figure}
The color plot shows the conductance of the device as a function of gate voltage and temperature after the first annealing step at 200 degree, but before the in situ annealing in the cryostat. At zero volts, the device is in the insulating regime. The transition to the metallic conduction can be anticipated at high voltages where conductance no longer increases with increasing temperature and also exhibits a value of about e$^2$/h, but the full metallic regime is not reached. The characteristics after the following in situ heating are shown in Figure 3 in the main text.\\
Figure 4 shows the characteristic annealing curves of the sample. The backgate voltage is kept at zero Volts. A permanent increase indicates an ongoing removal of adsorbents. During the cool down a further steep increase is observed, showing that the device is in the metallic regime. The small features are due to short changes in pressure and temperature.
\begin{figure}[ht]
\includegraphics[width=0.75\columnwidth]{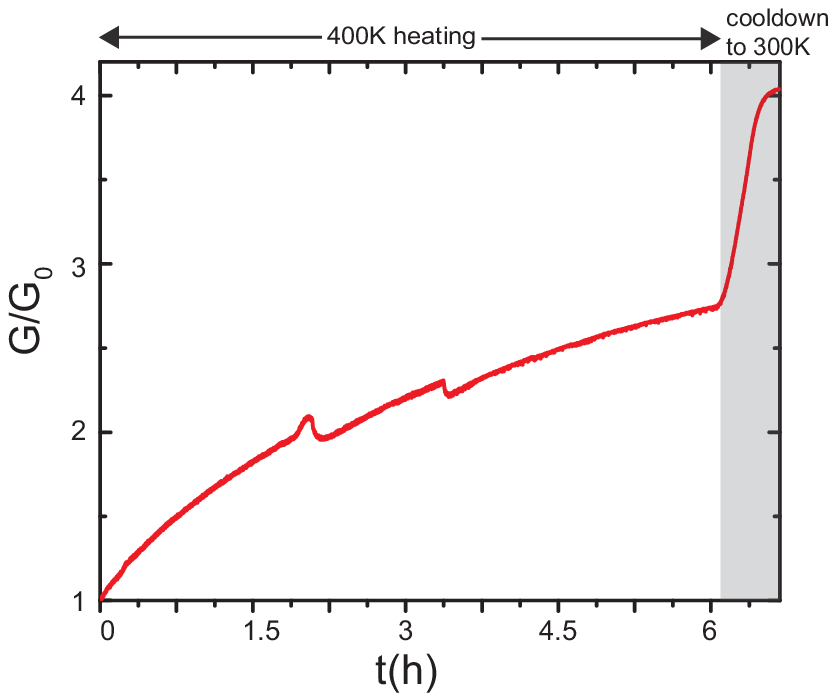}
\caption{\label{fig:bild4} Normalized conductance as a function of time during the in situ annealing in the cryostat. The grey shaded part indicated the cool down process}
\end{figure}
\newpage
\section{Multi-terminal measurements}
Four-terminal measurements are performed in a defined Hall-bar geometry with a constant current of 100~nA. As expected due to the absence of contact resistance, the field effect mobility yields higher values of around 500 cm$^2$V$^{-1}$s$^{-1}$ at low temperature. Figure 5 shows a multiterminal measurement on a device with a length to width ratio of 1.66. We observed that these measurements were much noisier than the two terminal ones.
\begin{figure}[ht]
\includegraphics[width=0.84\columnwidth]{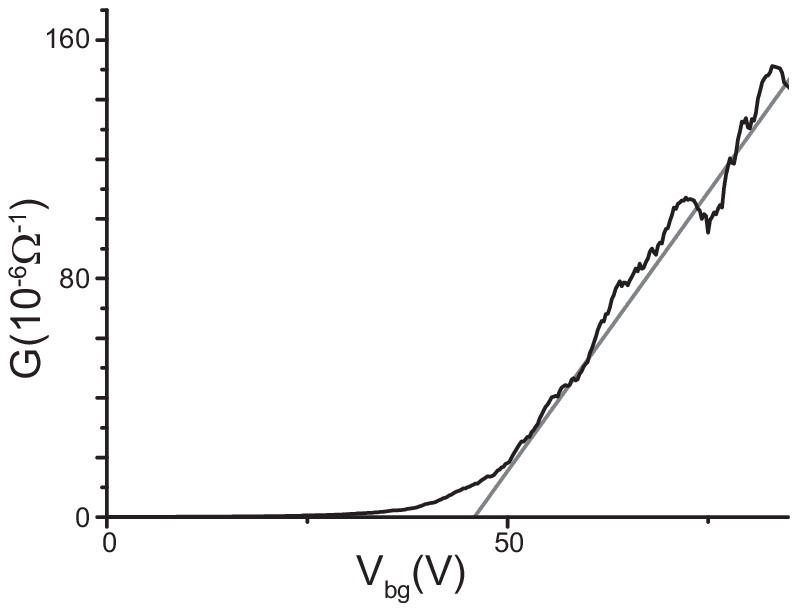}
\caption{\label{fig:bild5} Conductance obtained from a multiterminal measurement at 10 K. The line indicates a linear fit with a slope of 3.3 10$^{-6}\Omega^{-1}$V$^{-1}$ to estimate the mobility of the device.  }
\end{figure}
\newpage

\section{Insulating regime}
In the insulating regime (V$_{bg}<$0~V) the data can well fitted by the formula given by variable range hopping,
$$\sigma =\sigma_0(T)exp[(\frac{T_0}{T})^{\frac{1}{d+1}}]$$
with dimension d=2 and $\sigma_0\propto T^{0.8}$, according to the results published by other groups \cite{s5,s6,s7}.
The top graph shows the values at -105 V backgate voltage, the values at other negative voltages can be fitted comparable well. From the fits the characteristic parameter T$_0$ is extracted, showing the decrease and the same range of values as reported before \cite{s5}.
\begin{figure}[ht]
\includegraphics[width=0.69\columnwidth]{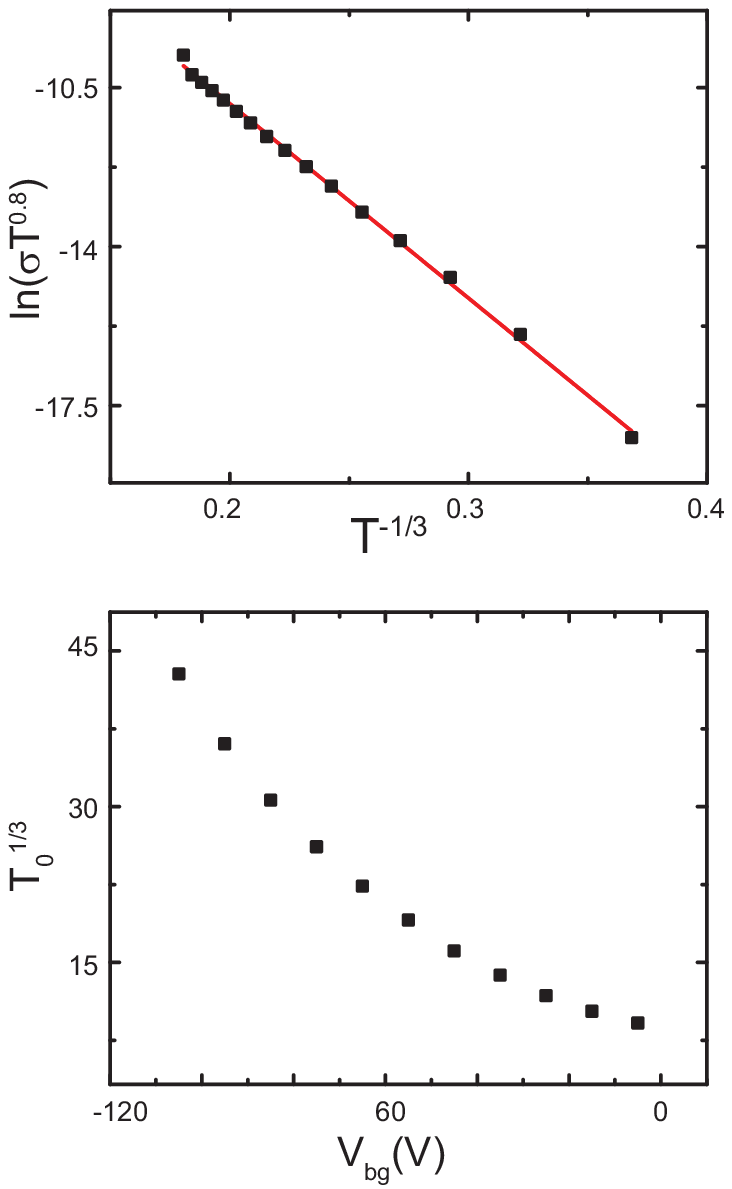}
\caption{\label{fig:bild6}Variable range hopping data and fit at a backgate voltage of -105 V. The bottom graph shows T$_0$ extracted from measurements at different voltages }
\end{figure}

\section{Metallic regime}
Figure 7 shows the results of another device not discussed in the main article. Conductance curves at different temperatures show that the crossing is further shifted, so the insulating regime is not fully reached at most negative voltages. The mobility shows the same behavior as depicted before, an increase with gate voltage, an overshoot at low temperatures at around -40 V and saturation at high gate voltages.\\
\begin{figure}[ht]
\includegraphics[width=0.69\columnwidth]{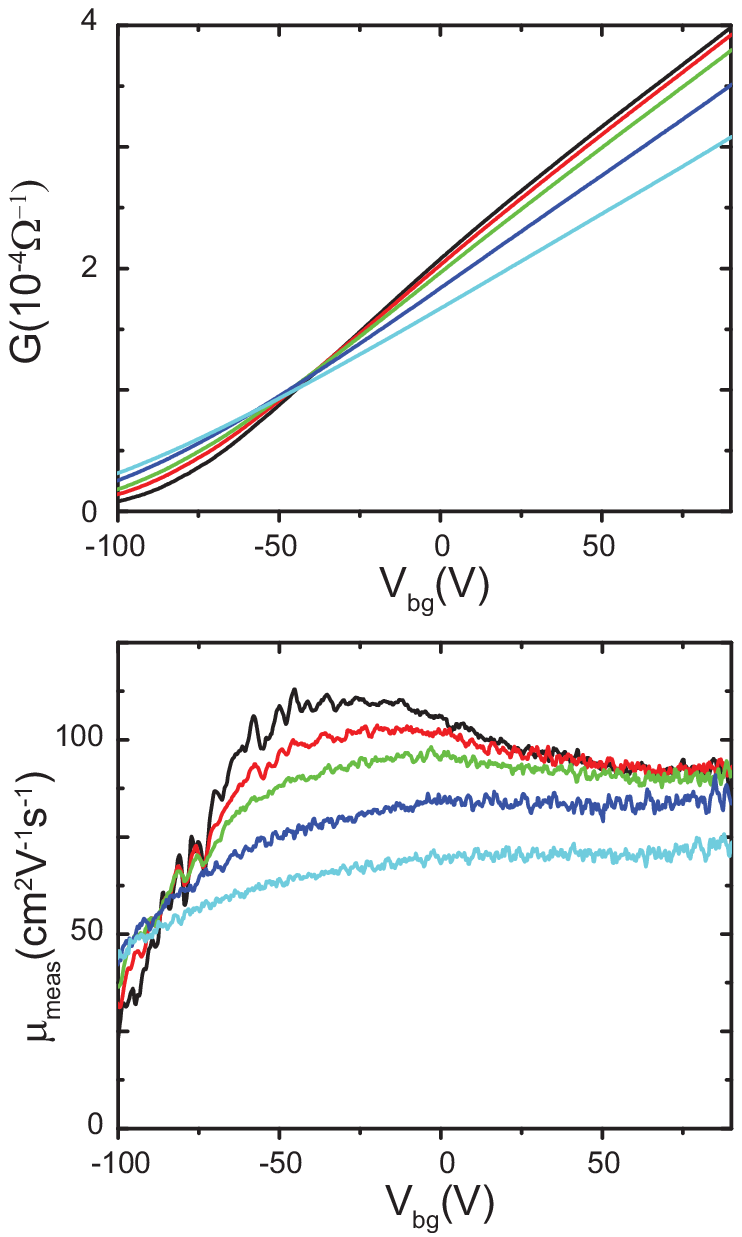}
\caption{\label{fig:bild7} Conductance curves of at 10, 40, 60, 100, and 150 K for a device with high doping level.  }
\end{figure}
A closer look at the region of crossing conductance curves at different temperatures at the transition from insulation to metallic behavior reveals that this crossing shifts with temperature (data from main text). The shift can be attributed to the thermal activation close to the conduction band edge, changing the amount of charge carriers with temperature. The right part of Figure 8 shows intersection points of the curves 5 K above and below of indicated temperature, e.g. the first point at 15 K indicates the crossing of the curves at 10 and 20 K. Extrapolating the quasi linear behavior indicated by the straight line gives a voltage of the zero temperature crossing at V$_{bg}$ = 16 V. The slope extracted from the linear fit exhibits a shift of -0.31 V/K.
\begin{figure}[ht]
\includegraphics[width=0.95\columnwidth]{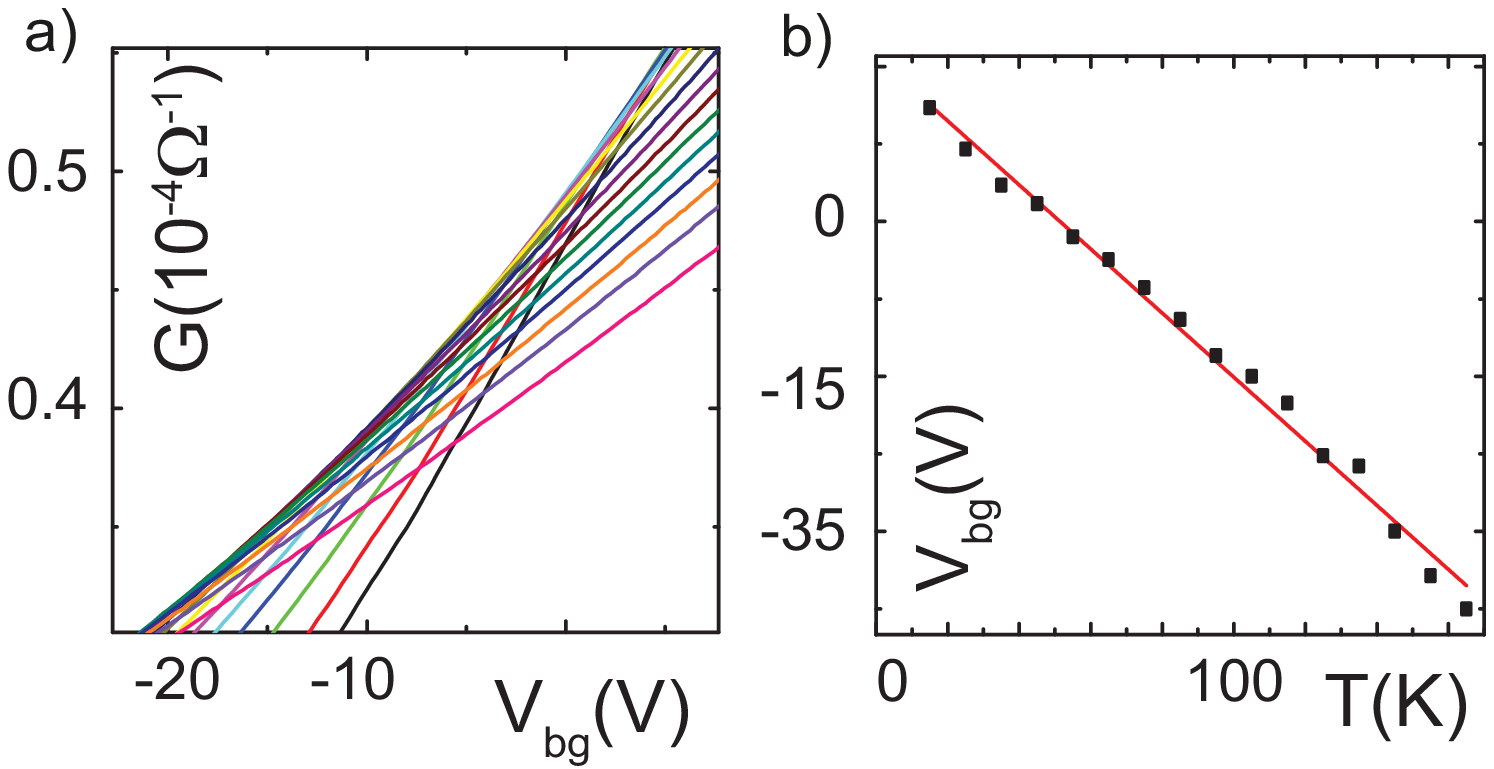}
\caption{\label{fig:bild8} Magnification of the crossing of conductance curves at different temperatures (10-170~K), taken from Figure 3b in the main text. Right graph shows the shift of crossing with temperature. }
\end{figure}
\newpage
\section{Temperature damping}
\begin{figure}[ht]
\includegraphics[width=0.95\columnwidth]{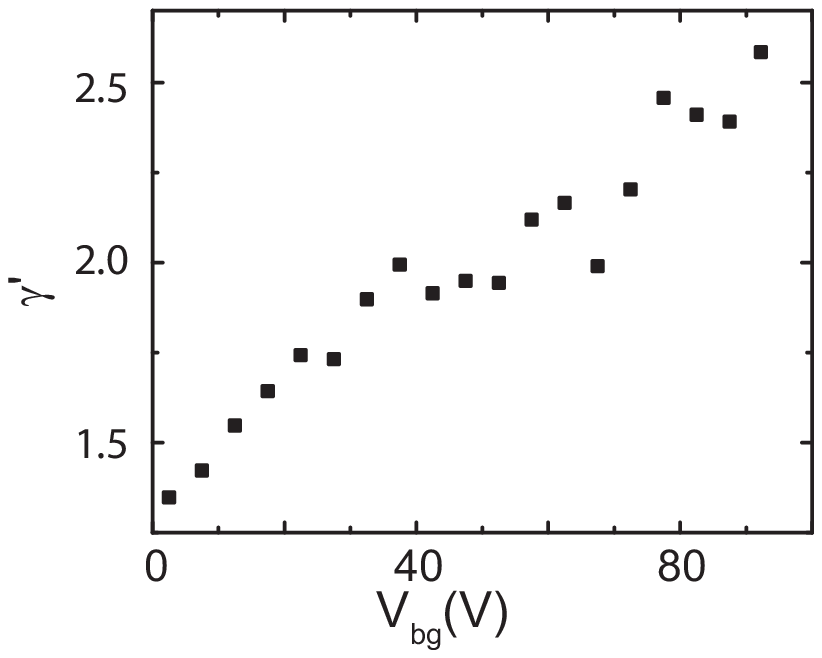}
\caption{\label{fig:bild9} The temperature damping factor as a function of back gate voltage.}
\end{figure}
The temperature damping factor is extracted from power law fits as shown in Figure 4d in the main text.  After the subtraction of the constant part attributed to short range scatterer, $\gamma`$ shows high values and strong gate voltage dependence (Fig. 9).
Figure 10 shows  $\mu`$ in different single layer devices grown by CVD in comparison with the data previously published data by Radisavljevic et al. \cite{s8} for exfoliated topgated samples, on which we did the same analysis. Same as for our devices, the fit suggests that mobilities of $> 10^3 cm^{2}V^{-1}s^{-1}$ could be reached if short range scatterers are fully removed from the material.
\begin{figure}[th]
\includegraphics[width=0.95\columnwidth]{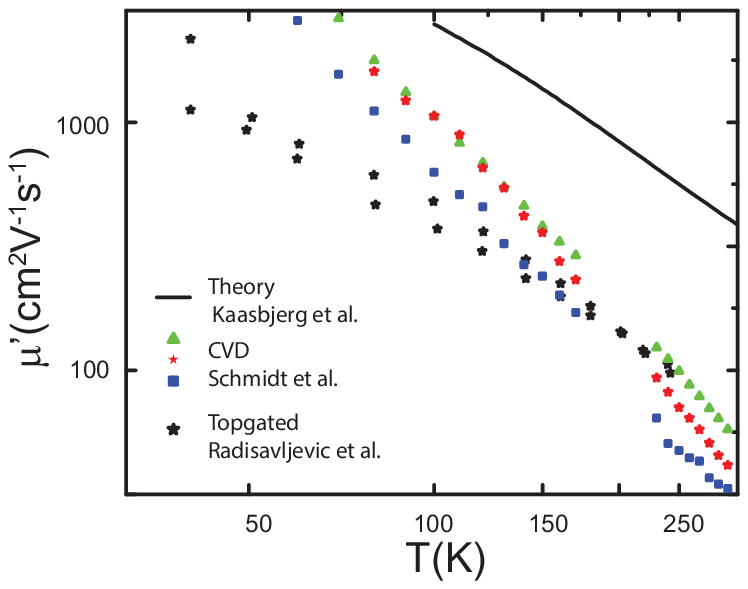}
\caption{\label{fig:bilds10}  $\mu`$ for three different devices in comparison to data from literature.}
\end{figure}

\newpage